\documentclass{revtex4}
\usepackage{amsmath}
\usepackage{amssymb}
\usepackage{graphicx}
\usepackage{ulem}
\begin{document}

\preprint{}

\title{Time--Evolving Statistics of Chaotic Orbits of Conservative Maps in the Context of the Central Limit Theorem}


\author{G. Ruiz}
\email[]{guiomar.ruiz@upm.es}
\affiliation{Departamento de Matem\'{a}tica Aplicada y Estad\'{\i}stica,
        Universidad Polit\'{e}cnica de Madrid, Pza. Cardenal Cisneros s/n, 28040 Madrid, Spain}
\author{T. Bountis}
\email[]{bountis@math.upatras.gr}
\affiliation{Department of Mathematics and Center for Research and Applications of Nonlinear Systems (CRANS) \\and University of Patras,
GR–26500, Rion, Patras (Greece)}
\author{C. Tsallis}
\email[]{tsallis@cbpf.br}
\affiliation{Centro Brasileiro de Pesquisas Fisicas and National Institute of Science and Technology for Complex Systems, Rua Xavier Sigaud 150, 22290-180 Rio de Janeiro, Brazil}
\affiliation{Santa Fe Institute, 1399 Hyde Park Road, Santa Fe, New Mexico 87501, USA}

\date{\today}

\begin{abstract}
We study chaotic orbits of conservative low--dimensional maps and present numerical results showing that the probability density functions (pdfs) of the sum of $N$ iterates in the large $N$ limit exhibit very interesting time-evolving statistics. In some cases where the chaotic layers are thin and the (positive) maximal Lyapunov exponent is small, long--lasting quasi--stationary states (QSS) are found, whose pdfs appear to converge to $q$--Gaussians associated with nonextensive statistical mechanics.  More generally, however, as $N$ increases, the pdfs describe a sequence of QSS that pass from a $q$--Gaussian to an exponential shape and ultimately tend to a true Gaussian, as orbits diffuse to larger chaotic domains and the phase space dynamics becomes more uniformly ergodic.
\end{abstract}


\maketitle

\section{Introduction}
As is well--known, invariant closed curves of area--preserving maps present complete barriers to orbits evolving inside resonance islands in the two--dimensional phase space. Outside these regions, there exist families of smaller islands and invariant Cantor sets (often called cantori), to which chaotic orbits are observed to ``stick'' for very long times. Thus, at the boundaries of these islands, an `edge of chaos' develops with vanishing (or very small) Lyapunov exponents, where trajectories yield quasi-stationary states (QSS) that are often very long--lived. Such phenomena have been thoroughly studied to date in terms of a number of \textit{dynamical} mechanisms responsible for chaotic transport in area--preserving maps and low--dimensional Hamiltonian systems (see e.g. \cite{MacKay, Wiggins}).

In this paper we study numerically the probability density functions (pdfs) of sums of iterates of QSS characterized by non--vanishing Lyapunov exponents, aiming to understand the connection between their intricate phase space dynamics and their time--evolving statistics. Our approach, therefore, is in the \textit{context} of the Central Limit Theorem (CLT), although in many cases our pdfs do not converge to a single shape but pass through several ones. One case where convergence is known to exist is when the dynamics is bounded and uniformly hyperbolic (as e.g. in the case of Sinai billiards) and the associated pdf is a Gaussian. However, even in nonhyperbolic conservative models, there are regions where trajectories are essentially ergodic and mixing, so that Gaussians are ultimately observed, as the number of iterations grows. In such cases the maximal Lyapunov exponent is positive and bounded away from zero. What happens, however, when the motion is ``weakly'' chaotic and explores domains with intricate invariant sets, where the maximal (positive) Lyapunov exponent is very small? It is the purpose of this work to explore the statistics of such regions and determine the type of QSS generated by their dynamics.

Recently, there has been a number of interesting studies of such pdfs of one--dimensional maps \cite{Tirnakli1,Tirnakli2,Gruiz,Afsar} and higher--dimensional conservative maps \cite{Queiros} in precisely `edge of chaos' domains, where the maximal Lyapunov exponent either vanishes or is very close to zero. These studies provide evidence for the existence of $q$-Gaussian distributions, in the context of the Central Limit Theorem. This sparked off fierce controversy \cite{Grassberger} but, for one--dimensional maps, the argument has been resolved. In fact, \cite{Tirnakli2, Tirnakli4} undoubtedly show that, when approaching the critical point while taking into account a proper scaling relation that involves the vicinity of the critical point and the Feigenbaum constant $\delta$, the pdfs of sums of iterates of the logistic map are approximated by a $q$-Gaussian far better that the L\'{e}vy distribution suggested in \cite{Grassberger}. This suggests the need for a more thorough investigation of these systems within a nonextensive statistical mechanics approach, based on the nonadditive entropy $S_q$ \cite{Tsallis,Tsallis2010}. According to this approach, the pdfs optimizing (under appropriate constraints) $S_q$ are $q$--Gaussian distributions that represent metastable states \cite{Miritello,Rodriguez,Baldovin1,Baldovin2}, or QSS of the dynamics.

The validity of a Central Limit Theorem (CLT) has been verified for deterministic systems \cite{Billingsley,Beck,Mackey} and, more recently, a $q$-generalization of the CLT was proved demonstrating that, for certain classes of strongly correlated random variables, their rescaled sums approach not a Gaussian, but a $q$-Gaussian limit distribution \cite{Umarov1, Umarov2,Hahn}. Systems statistically described by power-law probability distributions (a special case of which are $q$-Gaussians) are in fact so ubiquitous \cite{Schroeder}, that some authors claimed that the normalization technique of a type of data that characterizes the measurement device is {\it one} of the reasons of their occurrence \cite{Vignat}: This is the case of normalized and centered sums of data that exhibit elliptical symmetry, {\it but not necessarily} the case of the iterates of deterministic maps, as can be inferred by the verification of a classical CLT for the paradigmatic example of the fully chaotic logistic map.

In this paper, we follow this reasoning and compute first, in weakly chaotic domains of conservative maps, the pdf of the rescaled sum of $N$ iterates, in the large $N$ limit, and for many different initial conditions. We then connect our results with specific properties of the phase space dynamics of the maps and distinguish cases where the pdfs represent long--lived QSS described by $q$--Gaussians. We generally find that, as $N$ grows, these pdfs pass from a $q$--Gaussian to an exponential (having a triangular shape in our semi-log plots), ultimately tending to become true Gaussians, as ``stickiness'' to cantori apparently subsides in favor of more uniformly chaotic (or ergodic) motion.

In section II we begin our study by a detailed study of QSS, their pdfs and corresponding dynamics in two--dimensional Ikeda and MacMillan maps. In section III we briefly discuss analogous phenomena in $4$--dimensional conservative maps and end with our conclusions in section IV.
\section{Two--dimensional area--preserving maps}

Let us consider two--dimensional maps of  the form:

\begin{equation}
\label{map2d}
\begin{cases}
x_{n+1}=f(x_n, y_n) \\
y_{n+1}=g(x_n,y_n)
\end{cases}
\end{equation}
and treat their chaotic orbits as generators of
random variables. Even though this is not true for the iterates of a single orbit, we may still regard as random sequences those produced by many independently chosen initial conditions. In \cite{Mackey}, the well known CLT assumption about the independence of $N$ identically distributed random variables was replaced by a weaker property that essentially means asymptotic statistical independence. Thus, we may proceed to compute the generalized rescaled sums of their iterates $x_{i}$ in the context of the classical CLT (see \cite{Billingsley,Beck,Mackey}):
\begin{equation}
Z_N=N^{-\gamma} \sum_{i=1}^{N} \left(x_i-\langle x \rangle  \right)
\label{variable}
\end{equation}
where $\langle \cdots \rangle$ implies averaging over a large number of iterations $N$ {\it and} a large number of randomly chosen initial conditions $N_{ic}$. Due to the possible nonergodic and nonmixing behavior, averaging over initial conditions is an important ingredient of our approach.

For fully chaotic systems ($\gamma=1/2$), the distribution of \eqref{variable} in the limit ($N\rightarrow \infty$) is expected to be a Gaussian \cite{Mackey}. Alternatively, however, we may define the non--rescaled variable $z_N$
\begin{equation}
z_N= \sum_{i=1}^{N} \left[x_i-\langle x \rangle  \right]
\end{equation}
and analyze the probability density function (pdf) of $z_N$ normalized by its variance (so as to absorb the rescaling factor $N^{\gamma}$) as follows:

First, we construct the sums $S^{(j)}_N$ obtained from the addition of $N$ $x$-iterates $x_i \, \,  (i=0,\dots ,N)$ of the map \eqref{map2d}
\begin{equation}
\label{sum}
S^{(j)}_N=\sum_{i=0}^{N}x^{(j)}_i
\end{equation}
where $(j)$ represents the dependence of $S^{(j)}_N$ on the randomly chosen initial conditions $x^{(j)}_0$, with $j=1,2,...,N_{ic}$. Next, we focus on the centered and rescaled sums
\begin{equation}
\label{Normvariable}
s^{(j)}_N\equiv \left(S^{(j)}_N-\langle S^{(j)}_N \rangle \right)/\sigma_N=\left(
\sum_{i=0}^{N}x^{(j)}_i-\frac{1}{N_{ic}}\sum_{j=1}^{N_{ic}}\sum_{i=0}^{N}x^{(j)}_i
 \right)/\sigma_N
\end{equation}
where $\sigma_N$ is the standard deviation of the $S^{(j)}_N$
\begin{equation}
\sigma_N^2=\frac{1}{N_{ic}}\sum_{j=1}^{N_{ic}}\left(S^{(j)}_N-\langle S^{(j)}_N \rangle \right)^2=  \langle \left.S^{(j)}_N \right. ^2\rangle-\langle S^{(j)}_N\rangle^2
\end{equation}

Next, we estimate the pdf of $s_N^{(j)}$, plotting the histograms of $P(s_N^{(j)})$ for sufficiently small increments $\Delta s_N^{(j)}$($=0.05$ is used in all cases), so as to smoothen out fine details and check if they are well fitted by a $q$-Gaussian:
\begin{equation}\label{qGaussian}
P(s_N^{(j)})=P(0)\left( 1+\beta (q-1)(s_N^{(j)})^2 \right)^{\frac{1}{1-q}}
\end{equation}
where $q$ is the index of the nonadditive entropy $S_q$ and $\beta$ is a `inverse temperature' parameter. Note that as $q\rightarrow1$ this distribution tends to a Gaussian, i.e., $\lim_{q\rightarrow 1}P(s_N^{(j)})= P(0)e^{-\beta(s_N^{(j)})^2}$. From now on, we write $z/\sigma  \equiv  s_N^{(j)}$. We also remark that, due to the projection of the higher dimensional motion onto a single axis, the support of our distributions appears to consist of a dense set of values in $z/\sigma$, although we cannot analytically establish its continuum nature.
\subsection{The Ikeda map}

\begin{figure}
\includegraphics[height=6cm,angle=0,clip=]{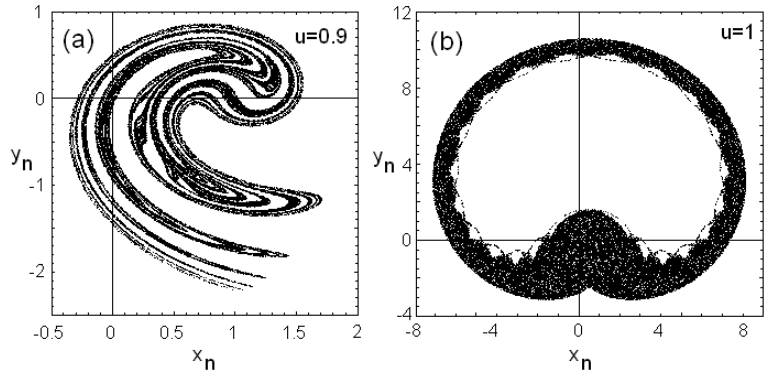}
\caption{Phase space plots of the Ikeda map for $C_1=0.4$, $C_2=6$, $R=1$, and representative values of $u$. When $u= 0.9$, areas of the phase plane contract and a strange attractors appears. When $u=1$, the map is area--preserving and a chaotic annular region is observed surrounding a domain about the origin where the motion is predominantly quasiperiodic. We use randomly chosen initial conditions from a square $[0,10^{-4}]\times [0,10^{-4}]$ about the origin (0,0).\label{fig_1}}
\end{figure}

Let us first examine by this approach the well--known Ikeda map \cite{Ikedapaper}:
\begin{equation}
\begin{cases}
x_{n+1}=R+u(x_n \cos{\tau}-y_n \sin{\tau})\\
y_{n+1}=u(x_n \sin{\tau}+y_n \cos{\tau})
\end{cases}
\label{Ikeda}
\end{equation}
where $\tau=C_1-C_2/(1+x_n^2+y_n^2)$, $R$, $u$, $C_1$, $C_2$ are free parameters, and the Jacobian is $J(R, u, \tau)=u^2$, so that \eqref{Ikeda} is dissipative for $u <1$ and area-preserving for $u=1$. This map was proposed as a model, under some simplifying assumptions, of the type of cell that might be used in an optical computer \cite{Ikedapaper}. Fixing the values of $C_1=0.4$, $C_2=6$ and $R=1$ we observe that when $u= 0.7,0.8,0.9$, areas of the phase plane contract and strange attractors appear. In Fig.~\ref{fig_1} we plot two different structures of the phase space dynamics for representative values of the parameter, $u$.
The values of the positive (largest) Lyapunov exponent $L_{max}$ in these cases are listed in the Table~\ref{Lyapikeda}.

\begin{table}
\caption{Maximal Lyapunov exponents of the Ikeda map, with $C_1=0.4$, $C_2=6$, $R=1$ and $u=0.7,0.8,0.9,1.0$.\label{Lyapikeda}}
 \begin{ruledtabular}
\begin{tabular}{ccccc}
$u $ & $ 0.7 $& $ 0.8 $ & $ 0.9 $ & $ 1.0 $ \\ \hline
$L_{max}$ & $0.334$ & $0.344$ & $0.5076$ & $0.118$  \\
\end{tabular}
\end{ruledtabular}
\end{table}

\begin{figure}\hspace{-0.5cm}
\includegraphics[height=6cm,angle=0,clip=]{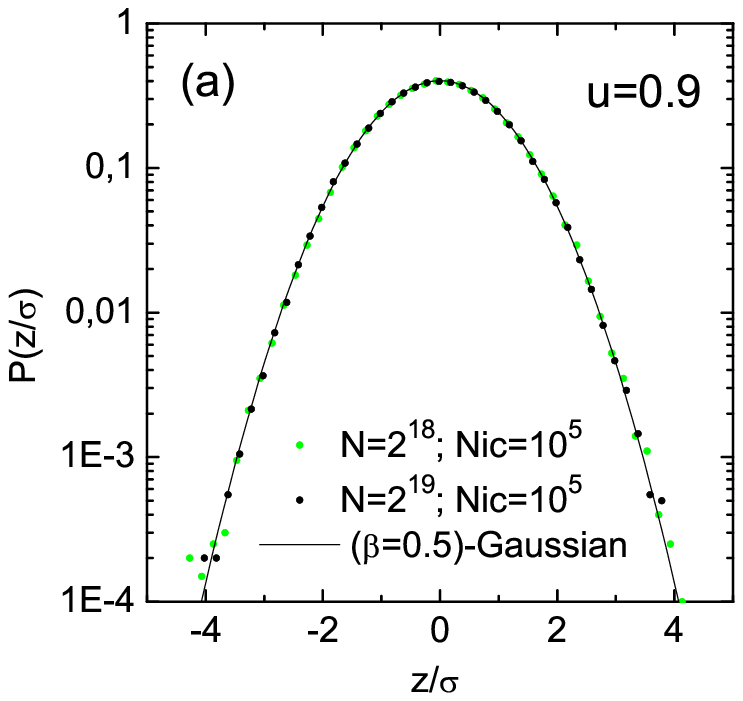}\hspace{-0.8cm}
\includegraphics[height=6cm,angle=0,clip=]{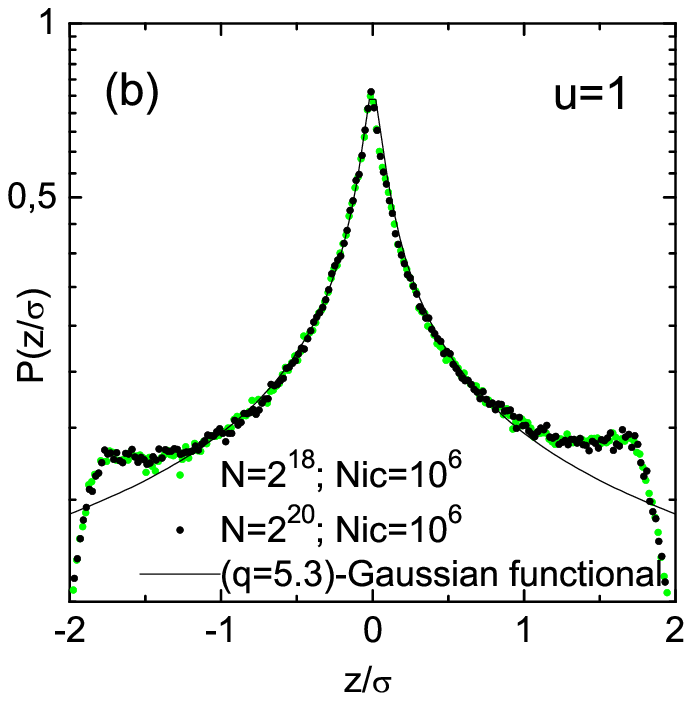}
\caption{(Color online) Pdfs of the normalized sums of iterates of the Ikeda map, for $C_1=0.4$, $C_2=6$, $R=1$. $N$ represents the number of (summed) iterates. Panel (a): $N_{ic}$ is the number of randomly chosen initial conditions from the basin of attraction (dissipative case); black line corresponds to Gaussian function $e^{-\beta(z/\sigma)^2}$, $\beta=0.5$. Panel(b): $N_{ic}$ is the number of randomly chosen initial conditions from a square $[0,1]\times [0,1]$ located inside the chaotic annular region of the area-preserving map; black line corresponds to $(q=5.3)$-Gaussian functional.  \label{fig_2}}
\end{figure}
Fig.~\ref{fig_2} shows the corresponding pdf of the normalized variables \eqref{Normvariable} obtained for the two values of the parameter, $u=0.9,1$, in the large $N$ limit. In fact, we observe that for $u=0.7,0.8,0.9$, the system possesses strange chaotic attractors whose pdfs can be properly fitted by Gaussians. These numerical results are not in disagreement with those of \cite{Tirnakli3}, on the $2$--dimensional H$\acute{e}$non map, where it was shown that its strange attractor exhibits nonextensive properties (i.e., $q\ne 1$). In a fully chaotic domain, non-extensive properties need not be present and consequently pdfs of the sum of iterates should be Gaussian distributions. Now, for  $u=0.7, 0.8, 0.9$, the Ikeda map \eqref{Ikeda} generates strange attractors whose maximum Lyapunov exponent is positive and bounded away from zero (see  Table~\ref{Lyapikeda}). This means that the motion is \textsl{not} at the `edge of chaos' but rather in a chaotic sea and consequently the concepts involved in Boltzmann-Gibbs statistics are expected to hold. On the contrary, in the area-preserving case $u=1$, the pdf of the sums of \eqref{Normvariable} converges to a non-Gaussian function (see Fig.~\ref{fig_2}b).

Now, in an `edge of chaos' regime, one might expect to obtain a $q$-Gaussian limit distribution ($q < 3$), which generalizes Gaussians and extremizes the nonadditive entropy $S_q$ \cite{Beck01} under appropriate constraints. Of course, the chaotic annulus shown in Fig.~\ref{fig_1} for $u=1$ does not represent an `edge of chaos' regime, as the maximal Lyapunov exponent does not vanish (see Table~\ref{Lyapikeda}) and the orbit appears to explore this annulus more or less uniformly. Hence a $q$-Gaussian distribution in that case would not be expected. But appearances can be deceiving. The result we obtain is remarkable, as the central part of our pdf is well--fitted by a $q$-Gaussian functional with $q=5.3$ up to very large $N$ (see Fig~.\ref{fig_2}b). Although this is not a normalizable $q$-Gaussian function (since $q > 3$ \cite{Tsallis2010}), it is nevertheless striking enough to suggest that: (a) the motion within the annular region is not as uniformly ergodic as one might have expected and (b) the $L_{max}$ is not large enough to completely preclude `edge of chaos' dynamics.

All this motivated us to investigate more carefully similar phenomena in another family of area-preserving maps.

\subsection{The MacMillan map}

Consider the so--called perturbed MacMillan map, which may be interpreted as describing the effect of a simple linear focusing system supplemented by a periodic sequence of thin nonlinear lenses \cite{Papageorgiou}:
\begin{equation}
\label{MacMillanpertb}
\begin{cases}
x_{n+1}=y_n\\
y_{n+1}=-x_n+2\mu\frac{y_n}{1+y_n^2}+\epsilon\left(y_n+\beta x_n\right)
\end{cases}
\end{equation}
where $\epsilon$, $\beta$, $\mu$ are physically important parameters. The Jacobian is $J(\epsilon, \beta)=1-\epsilon \beta$, so that \eqref{MacMillanpertb} is area-preserving for $\epsilon \beta=0$, and dissipative for $\epsilon\beta >0$. Here, we only consider the area-preserving  case $\beta=0$, so that the only relevant parameters are $(\epsilon$, $\mu)$.

The unperturbed map yields a lemniscate invariant curve with a self-intersection at the origin that is a fixed point of saddle type.
For $\epsilon \ne 0$, separatrices split and the map presents a thin chaotic layer around two islands. Increasing $\epsilon$, chaotic regions spread in the $x_n,y_n$ plane.

\begin{figure*}
   \includegraphics[width=16cm,angle=0]{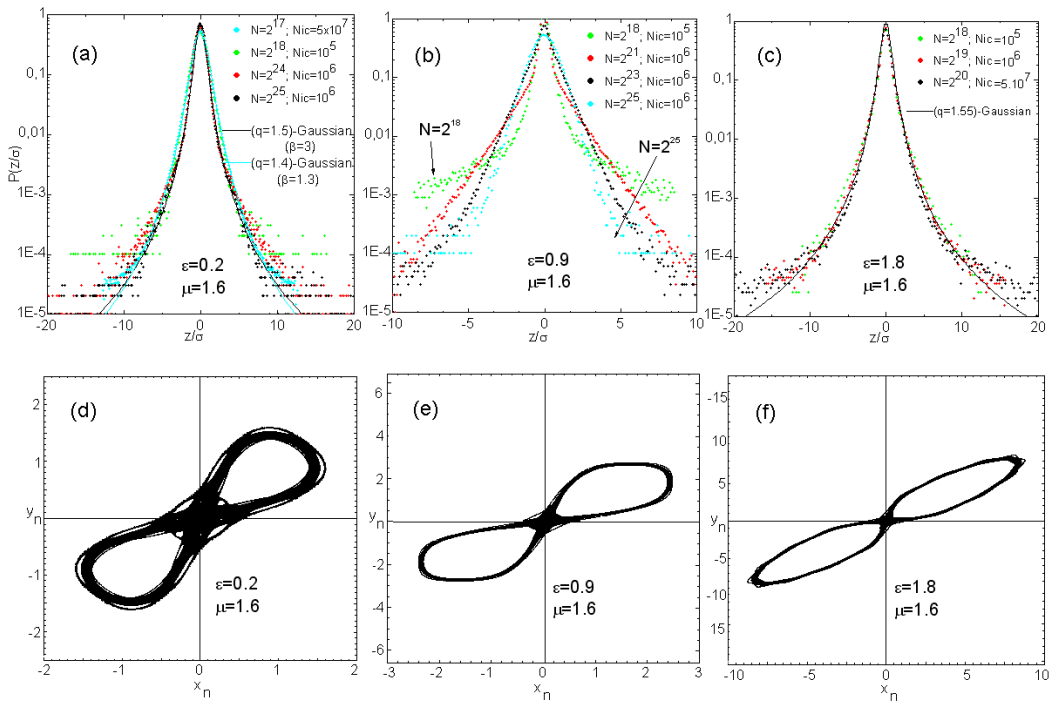}
\caption{(Color online) Dynamical and statistical behavior of chaotic orbits of the MacMillan map for parameter values $\mu=1.6$, and $\epsilon=0.2, 0.9, 1.8$ (from left to right). Figs.(a)-(c) represent the pdfs of the normalized sums of $N$ iterates; $N_{ic}$ is the number of randomly chosen initial conditions, from a square $(0,10^{-6})\times (0,10^{-6})$. Figs.(d)-(f) depict the corresponding phase space plots. \label{fig_3} }
\end{figure*}

\begin{figure*}
  \includegraphics[width=16cm,angle=0]{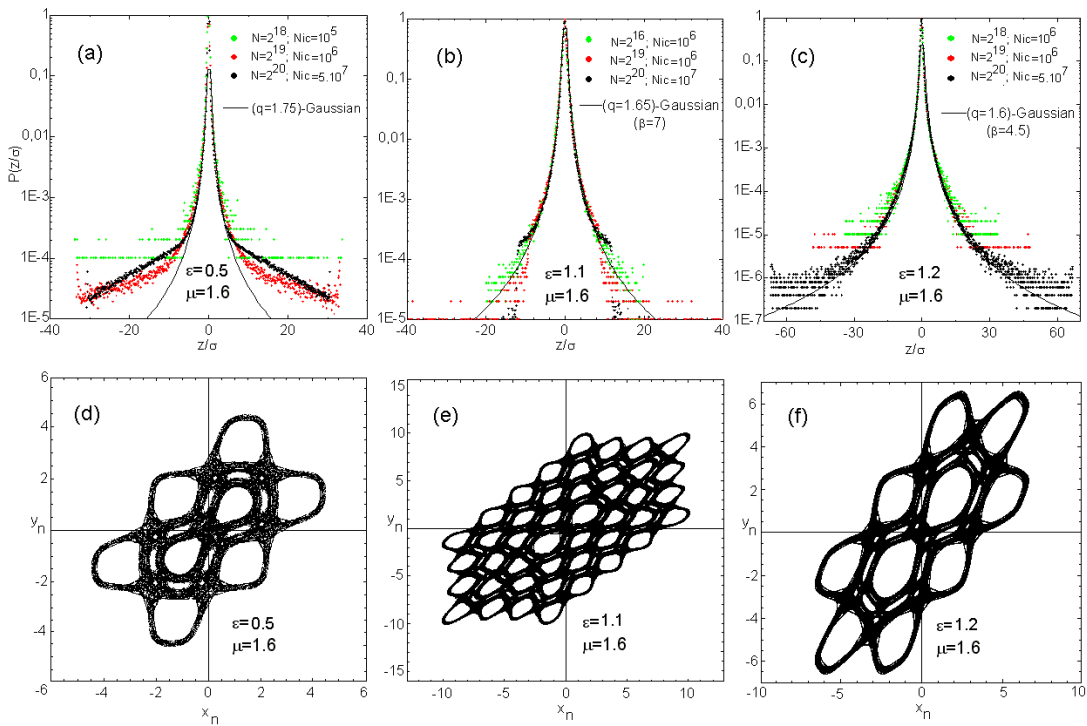}
\caption{(Color online) Dynamical and statistical behavior of chaotic orbits of the MacMillan map for parameter values $\mu=1.6$, and $\epsilon=0.5, 1.1, 1.2$ (from left to right), where the orbits evolve around a central `figure eight' chaotic region. Figs. (a)-(c) represent the pdfs of the normalized sums of $N$ iterates; $N_{ic}$ is the number of randomly chosen initial conditions, from a square $(0,10^{-6})\times (0,10^{-6})$. Figs. (d)-(f) depict the corresponding phase space plots. \label{fig_4} }
\end{figure*}

Within these chaotic regions, we have analyzed the histogram of the normalized sums of \eqref{Normvariable} for a wide range of parameters ($\epsilon$, $\mu$) and have identified some generic pdfs in the form of {\it $q$-Gaussians}, and {\it exponentials} $\sim e^{-k|z|}$ having a triangular shape on semi-logarithmic scale, which we call for convenience {\it triangular distributions}. Monitoring their `time evolution' under increasingly large numbers of iterations $N$, we typically observe the occurrence of \textit{different} QSS described by these distributions. We have also computed their $L_{max}$ and corresponding phase space plots and summarized our main results in Figures~\ref{fig_3} and \ref{fig_4}.
The maximal Lyapunov exponents for the cases shown in Figures \ref{fig_3} and \ref{fig_4} are listed in Table~\ref{LyapMacMillan}.

\begin{table}
\caption{Maximal Lyapunov exponents of the MacMillan map, with $\mu=0.6$ and $\epsilon=0.2,0.5,0.9,1.1,1.2,1.8$.\label{LyapMacMillan}}
 \begin{ruledtabular}
\begin{tabular}{ccccccc}
$\epsilon $ & $ 0.2$ & $ 0.5 $& $ 0.9 $  & $ 1.1 $ & $ 1.2 $ & $ 1.8$ \\ \hline
$L_{max}$ & $0.0867$ & $ 0.082$& $0.0875$ & $0.03446$& $0.0513$ & $0.05876$  \\
\end{tabular}
 \end{ruledtabular}
\end{table}

Below, we discuss the time-evolving statistics of two examples of the Mac Millan map, which represent respectively: (1) One set of cases with a `figure eight' chaotic domain whose distributions pass through a succession of pfds before converging to an ordinary Gaussian (Fig.~\ref{fig_3}), and (2) a set with more complicated chaotic domains extending around many islands, where $q$-Gaussian pdfs dominate the statistics for very long times and convergence to a Gaussian is not observed (Fig.~\ref{fig_4}).

\subsubsection{($\epsilon=0.9$, $\mu=1.6$)--MacMillan map}

The ($0.9$, $1.6$)--MacMillan map is a typical example producing time--evolving pdfs. As shown in Fig.~\ref{fig_3}, the corresponding phase space plots yield a seemingly simple chaotic region in the form of a `figure eight' around two islands, yet the corresponding pdfs do {\it not converge} to a single distribution; rather they pass from a $q$-Gaussian-looking function to a {\it triangular} distribution.

\begin{figure}
 \includegraphics[width=7cm,angle=0]{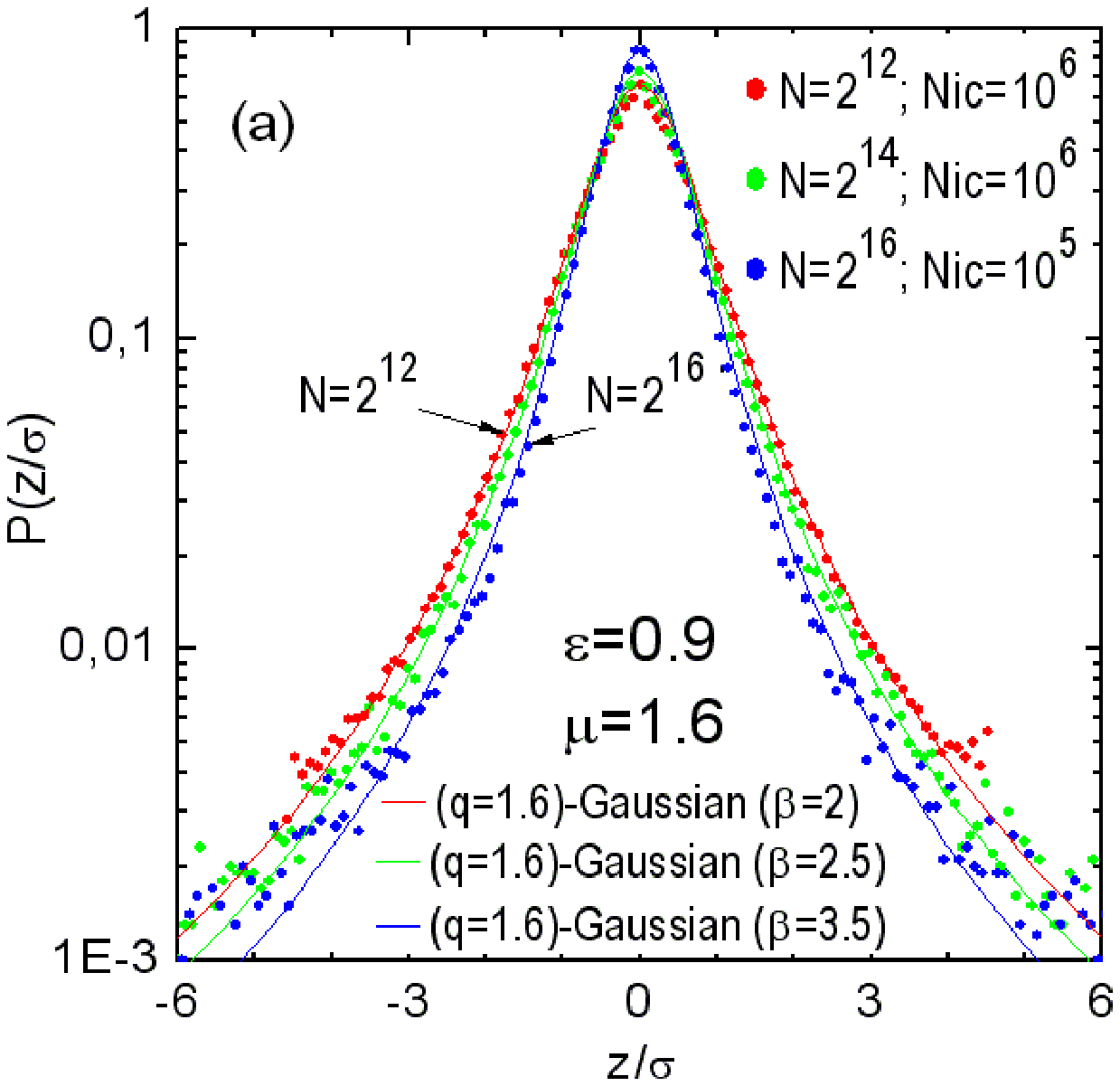}
  \includegraphics[width=6cm,angle=0]{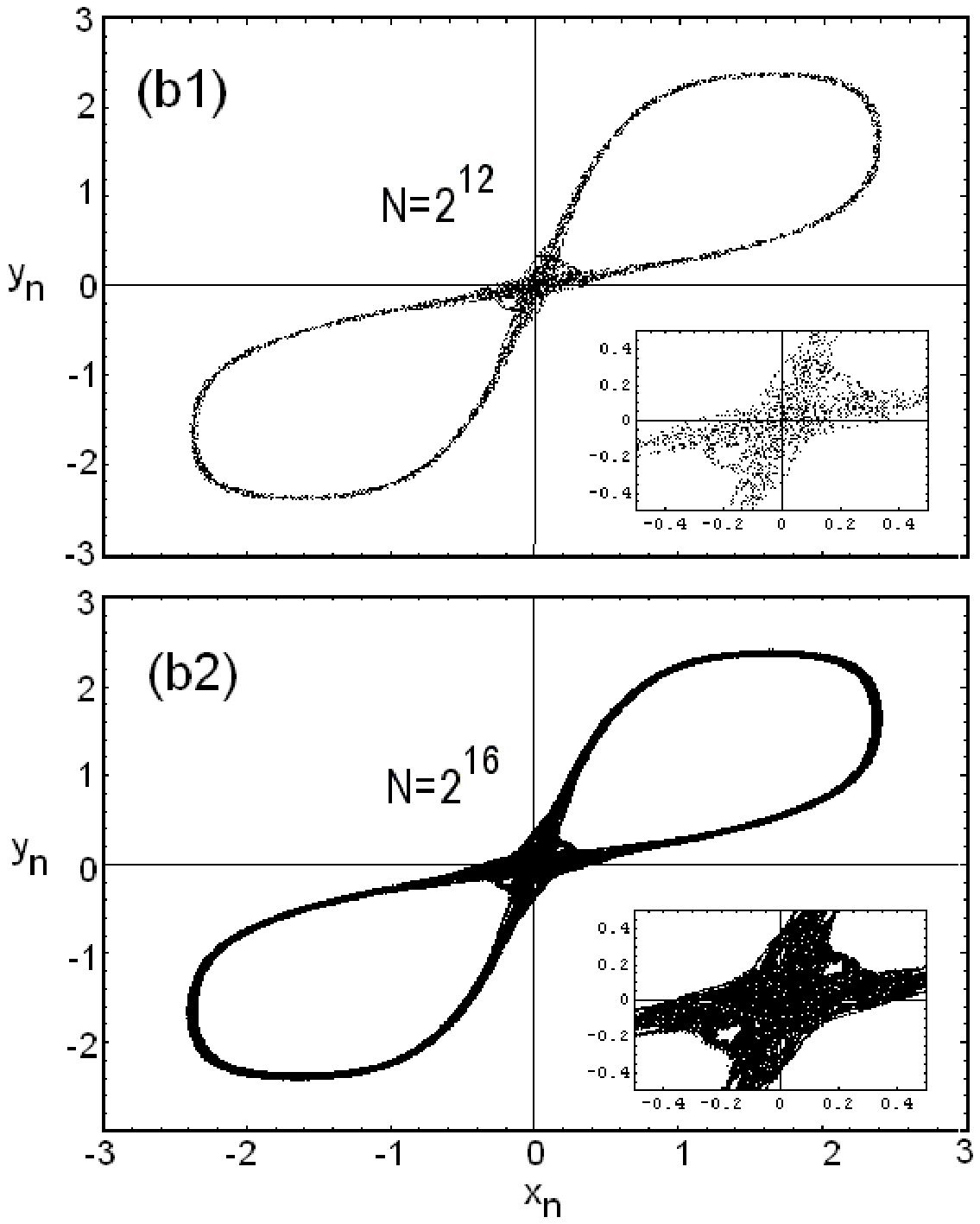}
  \caption{(Color online) Panel (a): PDFs of the renormalized sums of $N$ iterates of the ($\epsilon=0.9$, $\mu=1.6$)--MacMillan map, for $N\le 10^{16}$, and $N_{ic}$ randomly chosen initial condition in a square $(0,10^{-6})\times (0,10^{-6})$. Panel (b1)-(b2): Corresponding phase space plots for $N=2^{12}$ and $N=2^{16}$.\label{fig_5} }
\end{figure}

   \begin{figure}
 \includegraphics[width=9cm,angle=0]{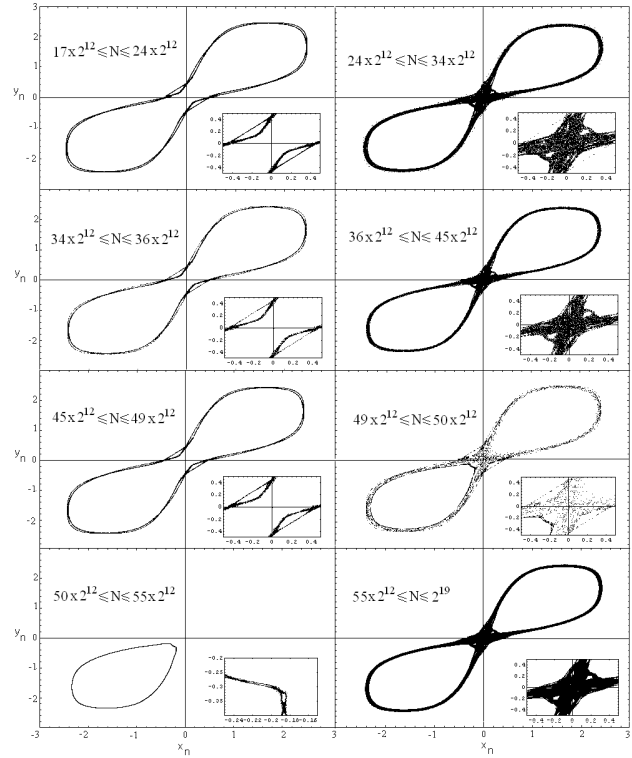}
 \caption{$(\epsilon=0.9$, $\mu=1.6$)--MacMillan map partial phase space evolution. The iterates are calculated starting form a randomly chosen initial condition in a square $(0,10^{-6})\times (0,10^{-6})$. $N$ is the number of plotted iterates. Note the long-standing quasi-stationary states that sequentially superimpose on phase space plots. \label{fig_6}}
  \end{figure}

    \begin{figure}\vspace{2cm}
  \includegraphics[width=7.6cm,angle=0]{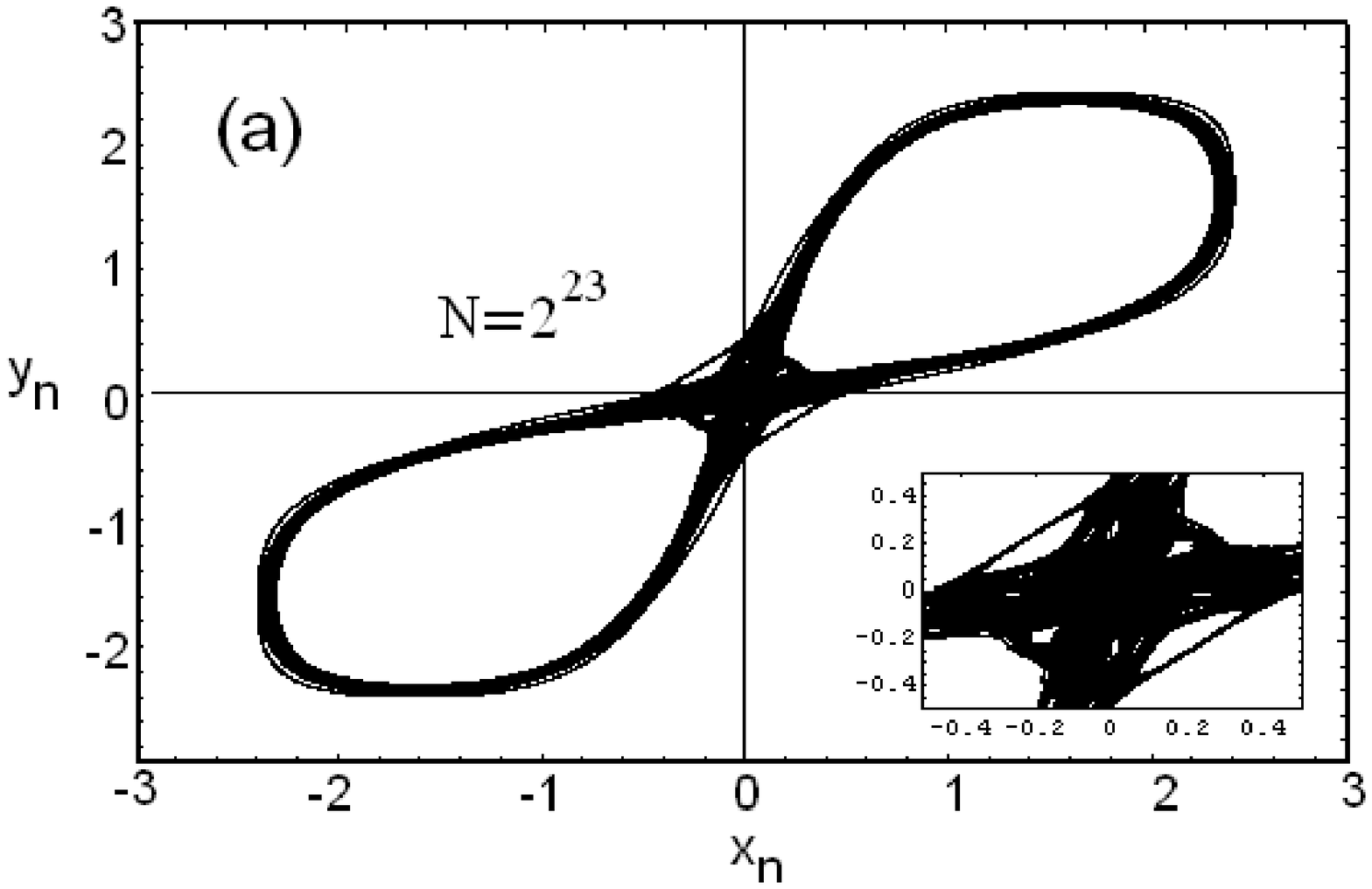}  \\\vspace{-4.4cm}
    \includegraphics[width=7.6cm,angle=0]{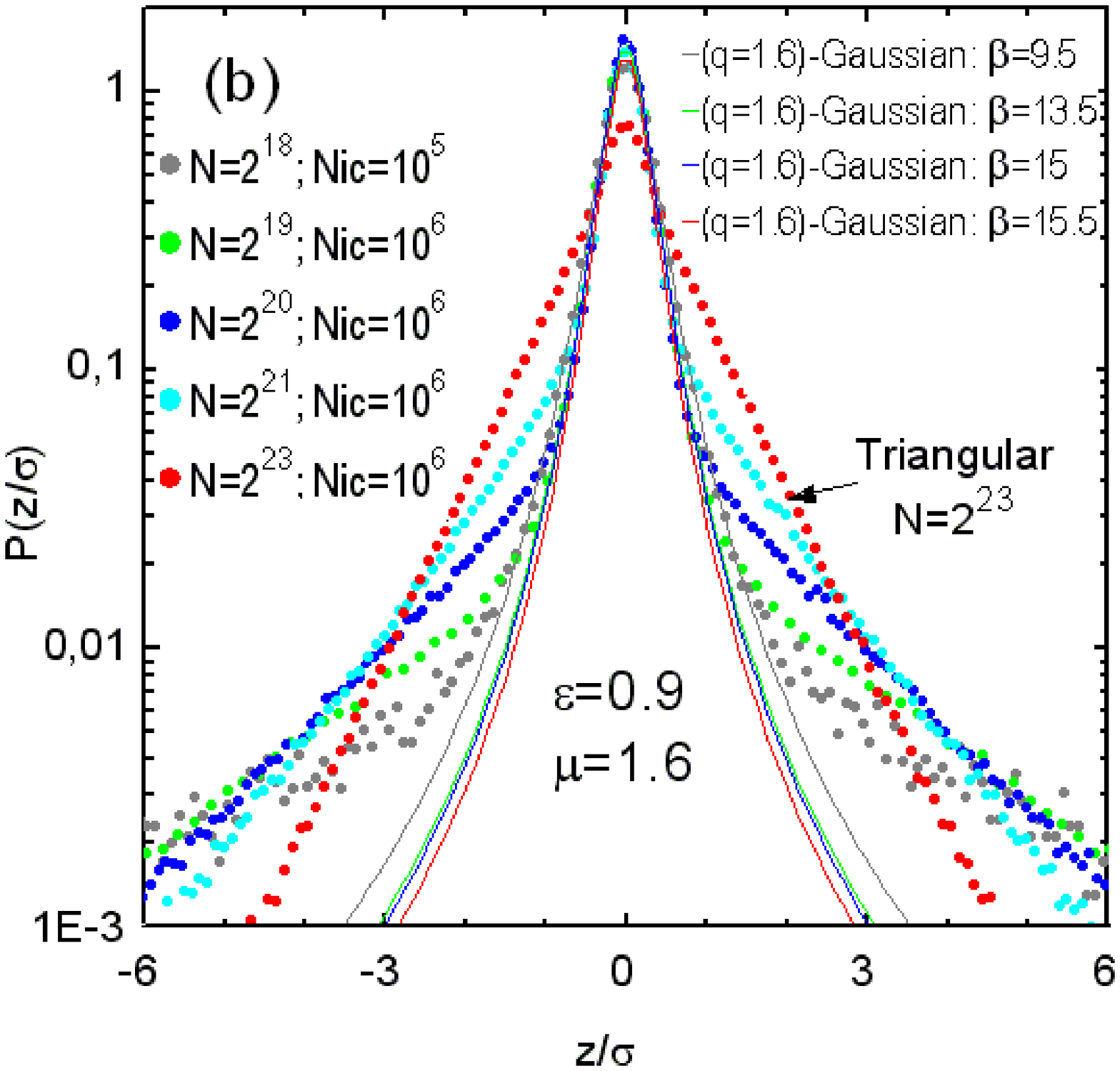}  \\\vspace{-4.5cm}
      \includegraphics[width=7.6cm,angle=0]{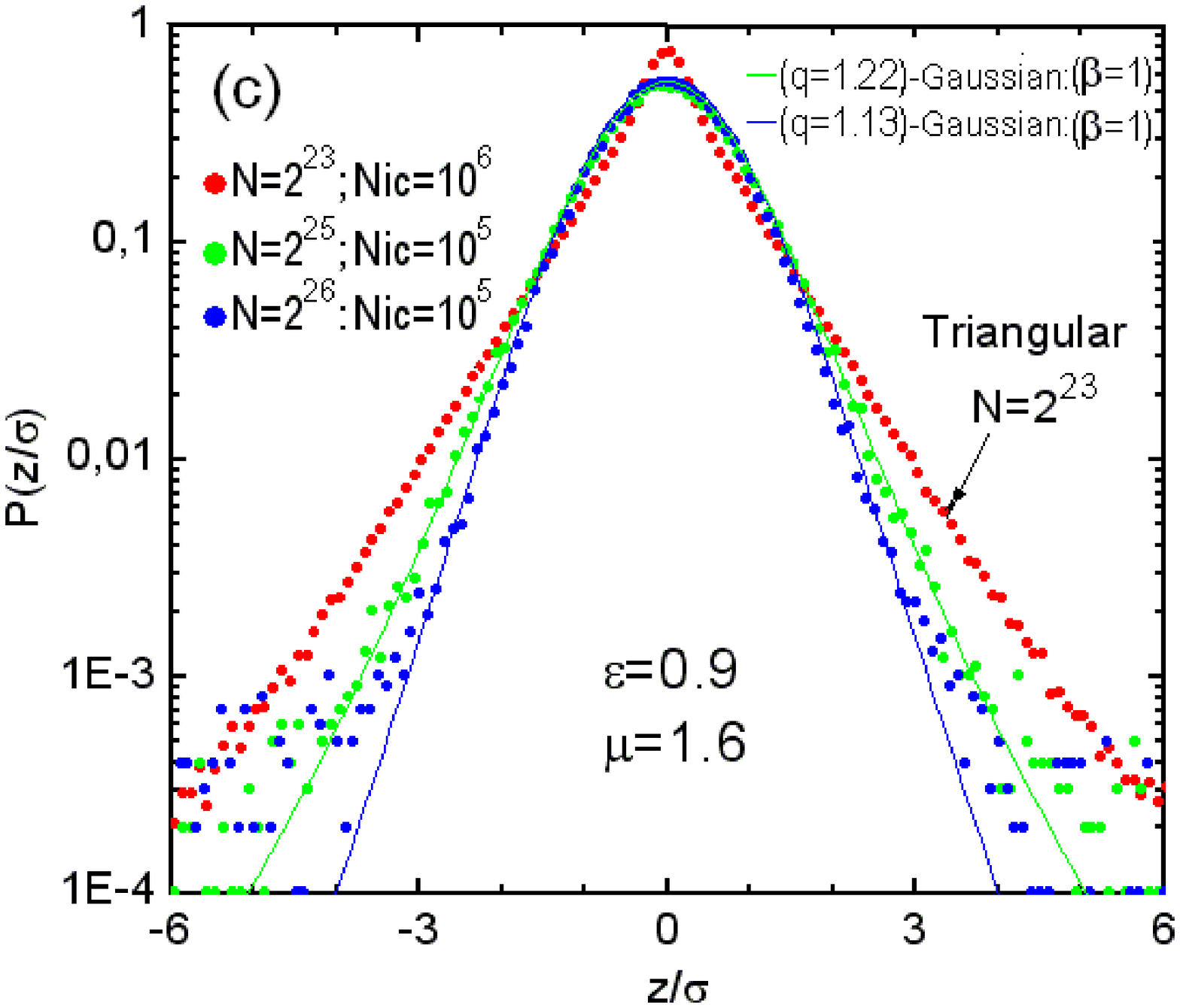}\vspace{-2cm}
 \caption{Panel (a): ($\epsilon=0.9$, $\mu=1.6$)--MacMillan  phase space plots for $i=1 \dots N\,  (N\ge 2^{23})$ iterates, starting form a randomly chosen initial condition in a square $(0,10^{-6})\times (0,10^{-6})$. Panel (b)-(c): (Color online) Corresponding PDFs. $N_{ic}$ is the number of randomly chosen initial condition in a square $(0,10^{-6})\times (0,10^{-6})$. \label{fig_7}}
\end{figure}

Analyzing carefully this time evolution of pdfs, we observed that there exist at least three long--lived QSS, whose iterates mix in the $2$--dimensional phase space to generate superimposed pdfs of the corresponding sums \eqref{Normvariable}. Consequently, for $i=1\dots N=2^{16}$, a QSS is produced whose pdf is close to a pure $(q=1.6)$--Gaussian whose $\beta$ parameter increases as $N$ increases and the density of phase space plot grows (see Fig.~\ref{fig_5}). This kind of distribution, in a fully chaotic region, is affected not only by a Lyapunov exponent being close to zero, but also by a ``stickiness'' effect around islands of regular motion. In fact, the boundaries of these islands is where the `edge of chaos' regime is expected to occur in conservative maps \cite{Weak}.

Fig.~\ref{fig_5} and Fig.~\ref{fig_6} show some phase space plots for different numbers of iterates $N$.  Note that for $N=1 \dots 2^{16}$, these plots depict first a `figure eight' chaotic region that evolves essentially around two islands (Fig.~\ref{fig_5}). However, for $N>2^{16}$, a more complex structure emerges: Iterates stick around new islands, and an alternation of QSS is evident from $q$-Gaussian to exponentially decaying shapes (see Fig.~\ref{fig_6}).

\begin{figure}
\hspace{1cm} \includegraphics[width=6cm,angle=0]{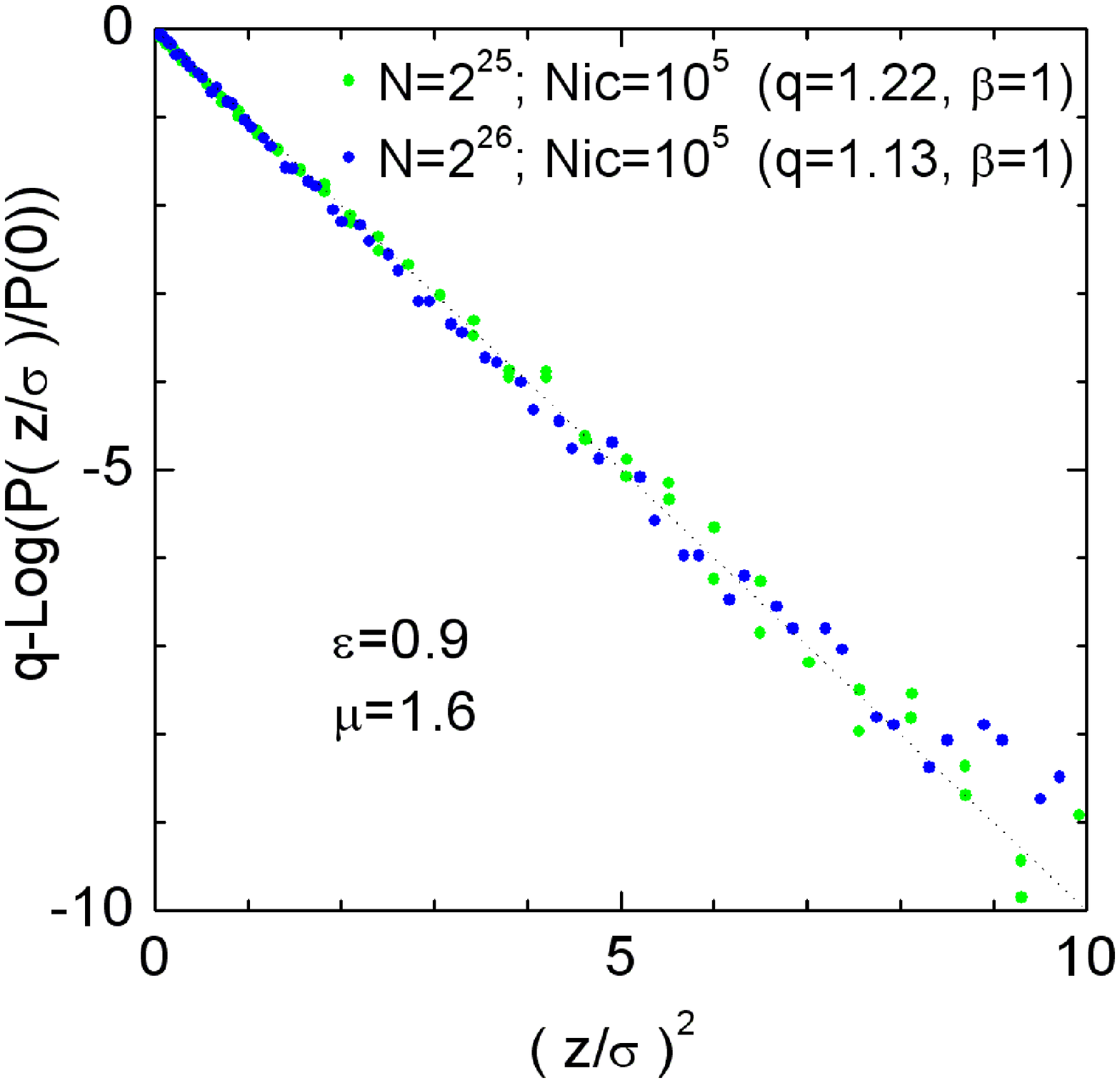}\vspace{-1cm}
\caption{(Color online) Plots of the $q$--logarithm (inverse function of the $q$-exponential \eqref{qGaussian}) vs. $(z/\sigma)^{2}$ applied to our data of the normalized pdf of the ($\epsilon=0.9$, $\mu=1.6$)--MacMillan map. $N$ is the number of iterates, starting from $N_{ic}$  randomly chosen initial condition in a square $(0,10^{-6})\times (0,10^{-6})$. For $q$--Gaussians this graph is a straight line, whose slope is $-\beta$) for the right value of $q$. Note that the pdfs approach a true Gaussian (with $\beta=1$) since $q$ tends to 1 as $N$ increases.\label{fig_8}}
\end{figure}

\begin{figure*}
\includegraphics[width=17.3cm,angle=0]{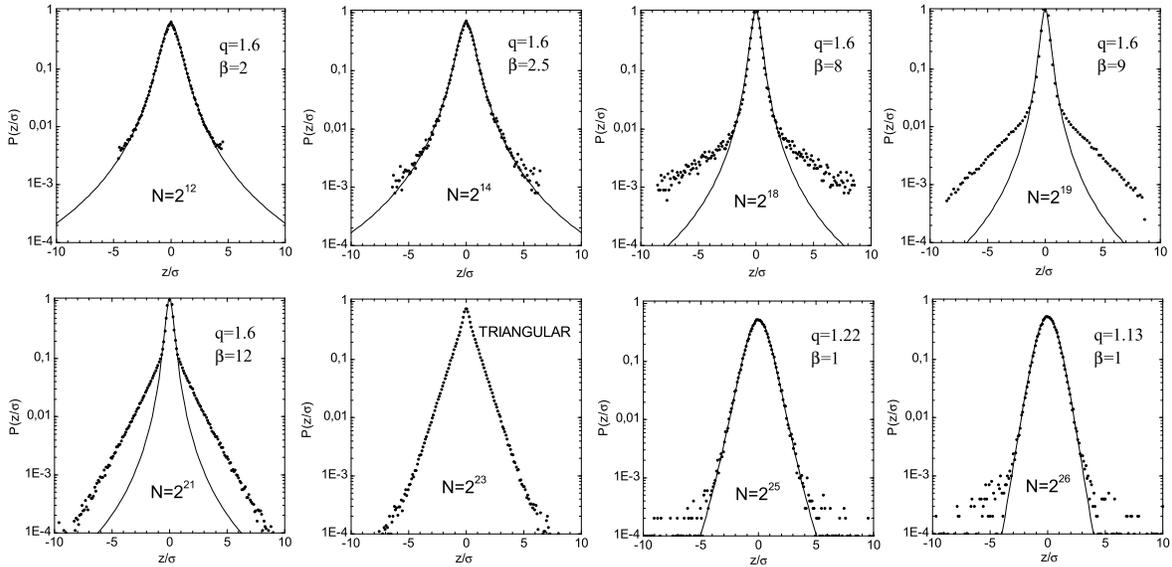}
\caption{Detailed evolution of the pdfs of  the MacMillan map for $\epsilon=0.9$, $\mu=1.6$, as $N$ increases from $2^{12}$ to $2^{26}$, respectively. \label{fig_9}}
\end{figure*}

Clearly, therefore, for $\epsilon=0.9$ (and other similar cases with $\epsilon=0.2,1.8$) more than one QSS coexist whose pdfs are the superposition of their corresponding $(q\ne 1)$--Gaussians. Note in Fig.~\ref{fig_7} that this superposition of QSS occurs for $10^{18}\le N \le 2^{21}$ and produces a mixed distribution where the central part is still well--described by a $(q=1.6)$--Gaussian. However, as we continue to iterate the map to $N =2^{23}$, this $q$--Gaussian is hidden by a superposition of intermediate states, passing to a {\it triangular} distribution. From here on, as $N > 2^{23}$, the central part of the pdfs is close to a Gaussian (see Fig.~\ref{fig_8} and Fig.~\ref{fig_9}) and a true Gaussian is expected in the limit ($N \rightarrow \infty$). The evolution of this sequence of successive QSS as $N$ increases is shown in detail in Fig.~\ref{fig_9}.

\subsubsection{($\epsilon=1.2$, $\mu=1.6$)--MacMillan map}

\begin{figure}
\hspace{1cm}  \includegraphics[width=7.5cm,angle=0]{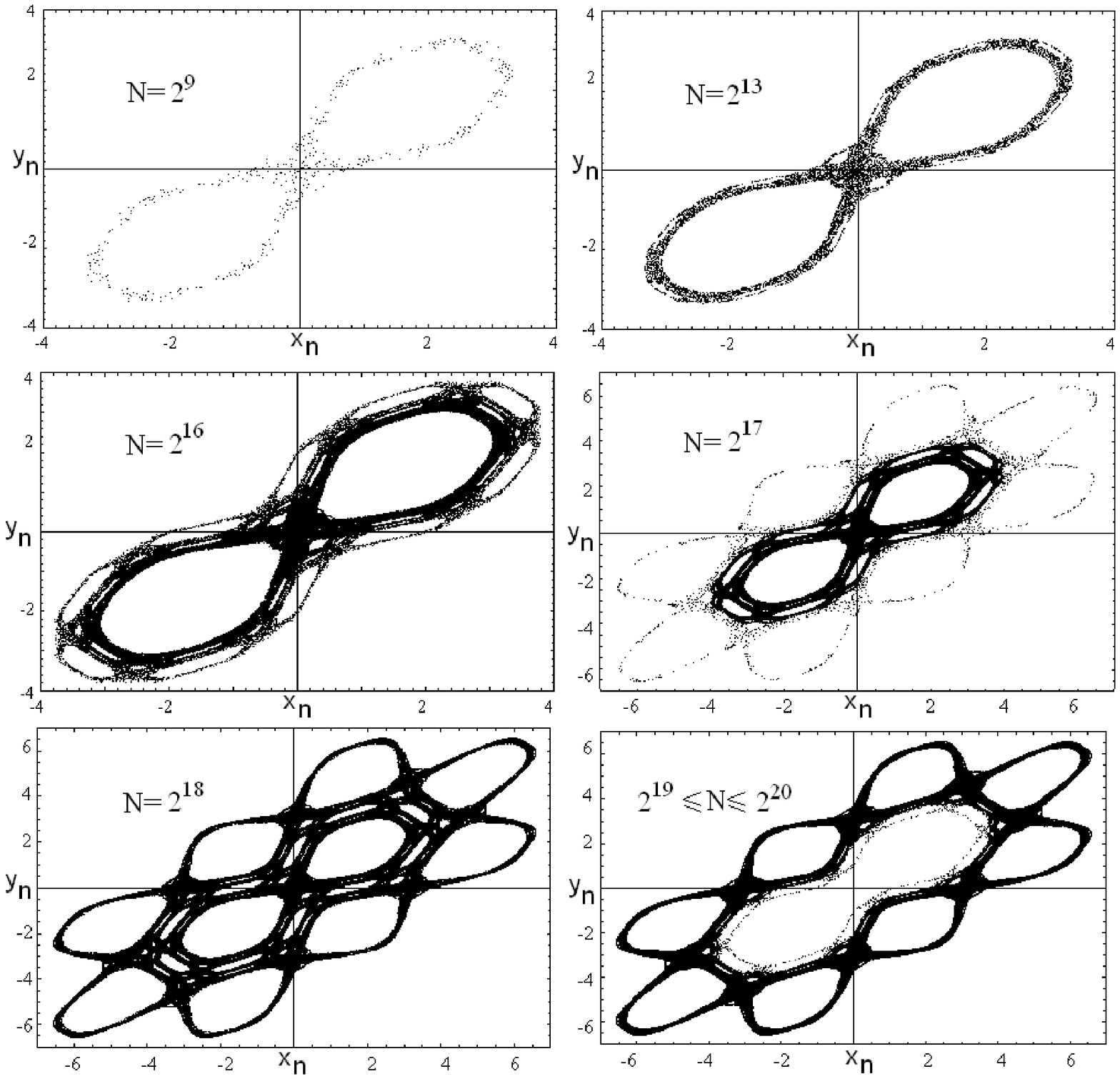}
    \caption{Structure of phase space plots of the MacMillan map for parameter values $\epsilon= 1.2$ and $\mu=1.6$, starting from a randomly chosen initial condition in a square $(0,10^{-6})\times (0,10^{-6})$, and for $ N$  iterates. \label{fig_10} }
\end{figure}

Let us now carefully analyze the behavior of the $(1.2, 1.6)$--MacMillan map, whose maximum Lyapunov exponent is $L_{max}\approx 0.05$, smaller than that of the $\epsilon=0.9$ case ($L_{max}\approx 0.08$). As is clearly seen in Fig.~\ref{fig_10}, a diffusive behavior sets in here that extends outward in phase space, envelopping a chain of islands of an order 8 resonance to which the orbits ``stick'' as the number of iterations grows to $N=2^{19}$.

\begin{figure}\vspace{-2.5cm}
 \includegraphics[width=8.5cm,angle=0]{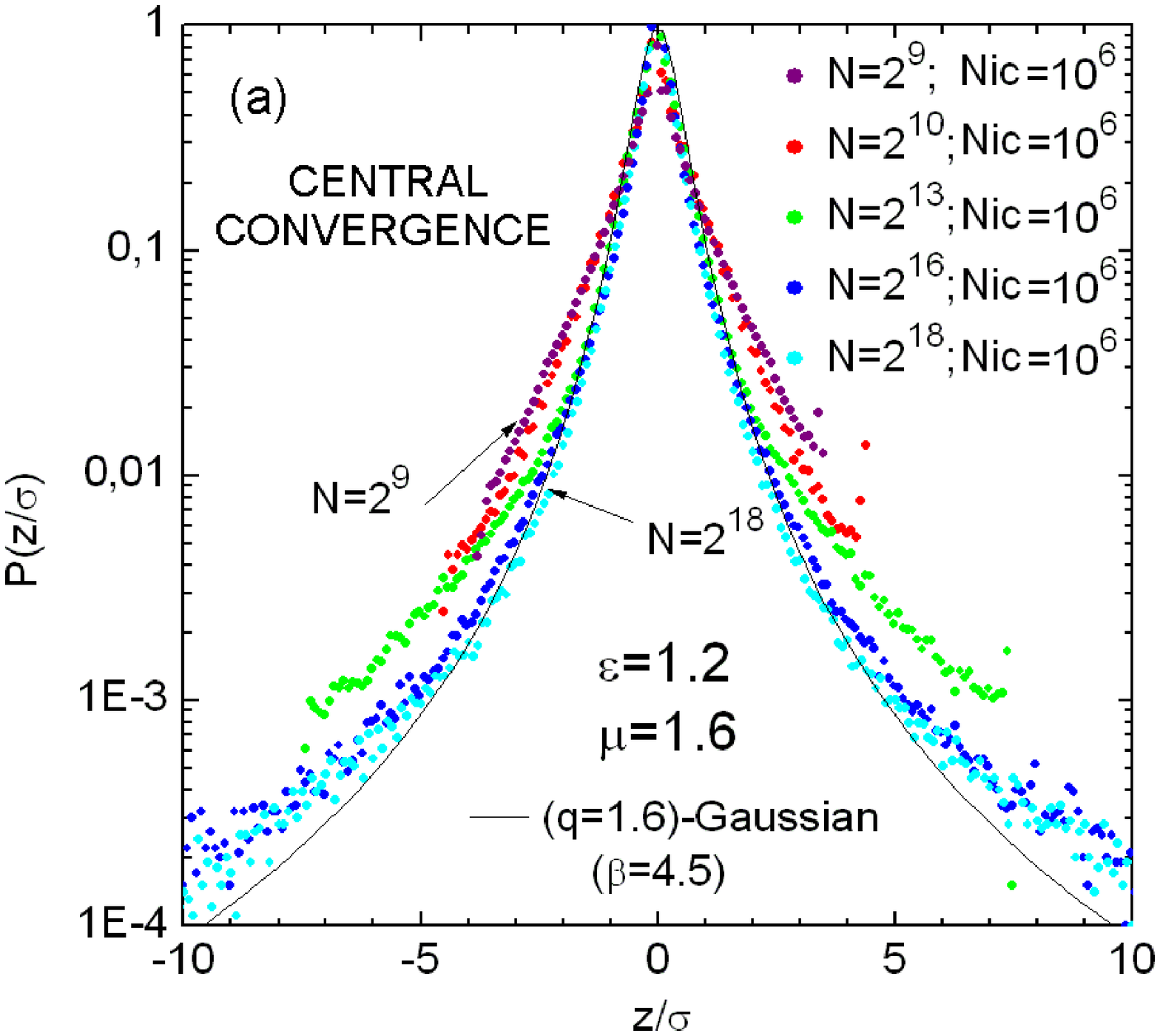}
  \includegraphics[width=8.5cm,angle=0]{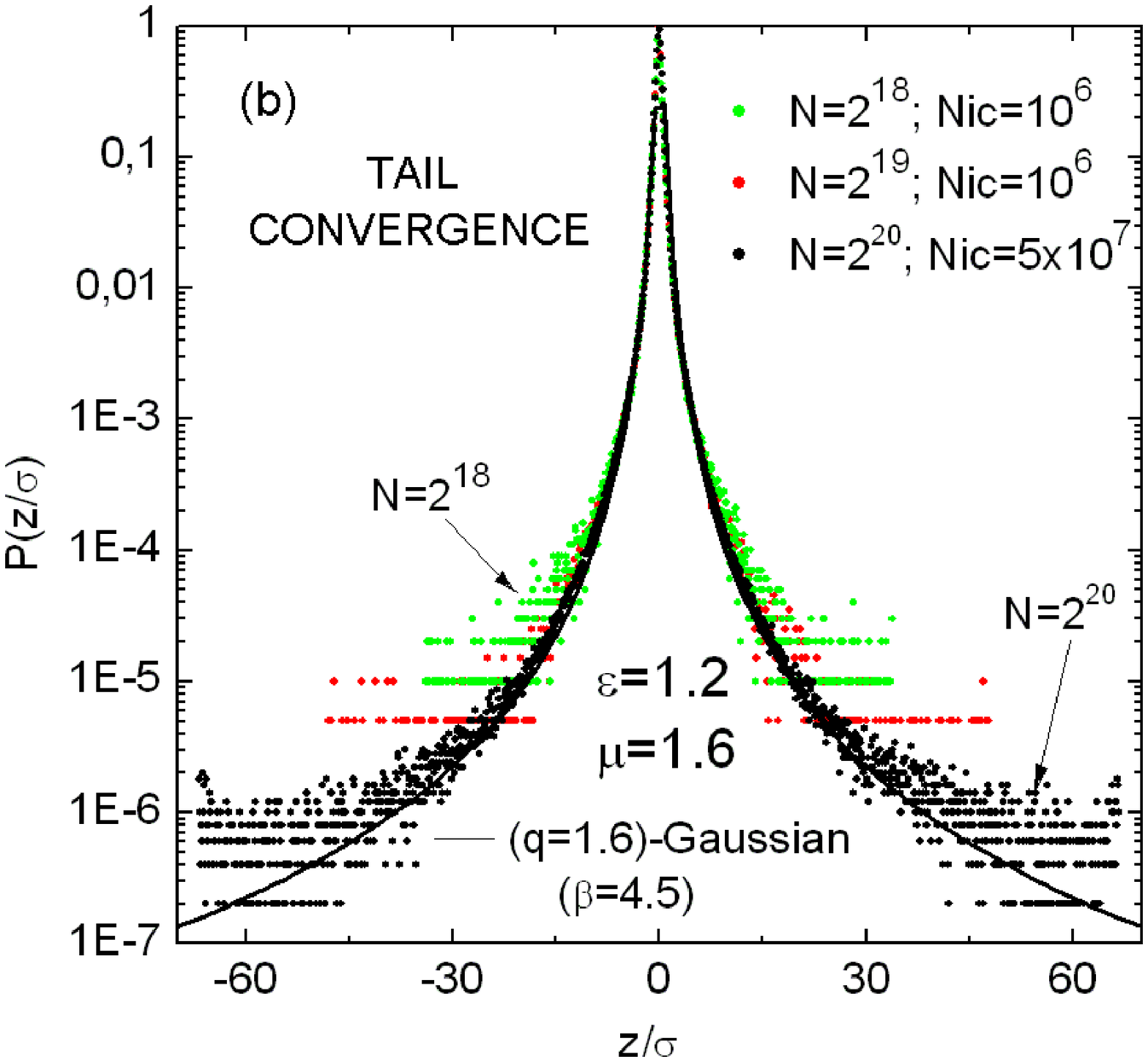}\vspace{-1cm}
  \caption{(Color online) Pdfs of the rescaled sums of iterates of the MacMillan map for $\epsilon= 1.2$ and $\mu=1.6$ are seen to converge to a $(q=1.6)$-Gaussian. This is shown in the panel (a) for the central part of the pdf (for $N<2^{18}$) and in the panel (b) for the tail part ($N>2^{18}$). $N_{ic}$ is the number of initial conditions that have been randomly chosen from a square $(0,10^{-6})\times (0,10^{-6})$\label{fig_11}}
\end{figure}

Again, chaotic motion starts by enveloping the same `figure eight' as in the $\epsilon=0.9$ case and the central part of the corresponding pdf attains a $(q=1.6)$-{\it Gaussian} form for $N \le 2^{16}$ (see Fig.~\ref{fig_11}a). No transition to a different QSS is detected, until the orbits diffuse to a wider chaotic region in phase space, for $N\le 2^{18}$. Let us observe in Fig.~\ref{fig_11}, the corresponding pdfs of the rescaled sums of iterates, where even the tail of the pdf appears to converge to a $(q=1.6)$-Gaussian (Fig.~\ref{fig_11}b). For larger $N$, further diffusion ceases as orbits ``stick'' to the outer islands, where the motion stays from there on. This only affects the tail of the distribution, which now further converges to a true $(q=1.6)$-{\it Gaussian} representing this QSS up to $N= 2^{20})$.

The remaining cases of Figures~\ref{fig_3} and \ref{fig_4} can be viewed from a similar perspective. Indeed, the above analysis of the $\epsilon=1.2$ example can serve as a guide for the $(\epsilon=0.5, \mu=1.6$)-- and $(\epsilon=1.1, \mu=1.6$)--MacMillan maps, as well. In every case, the smallness of the $L_{max}$ but also the details of the diffusion process seem to play a key role in explaining the convergence of pdfs to a $q$-Gaussian. What differs is the particular phase space picture that emerges and the number of iterations required to achieve the corresponding QSS.

We conclude, therefore, that the dynamics of the MacMillan map for $\mu=1.6$ and $\epsilon=0.2, 0.9, 1.8$, where chaotic orbits evolve around the two islands of a single `figure eight' chaotic region possess pdfs which pass rather quickly from a $q$-Gaussian shape to exponential to Gaussian. By contrast, the cases with  $\epsilon=0.5, 1.1, 1.2$ possess a chaotic domain that is considerably more convoluted around many more islands and hence apparently richer in ``stickiness'' phenomena. This higher complexity of the dynamics may very well be the reason why these latter examples have QSS with $q$--Gaussian-like distributions that persist for very long. Even though we are not at an `edge of chaos' regime where $L_{max}=0$, we suggest that it is the detailed structure of chaotic regions, with their network of islands and invariant sets of cantori, which is responsible for obtaining QSS with long-lived $q$-Gaussian distributions in these systems.

\section{Four--dimensional conservative maps}

We now briefly discuss some preliminary results on the occurrence of QSS and nonextensive statistics in a $4$--dimensional symplectic mapping model of accelerator dynamics \cite{BountisKollmann}. This model describes the effects of sextupole nonlinearities on a hadron beam passing through a cell composed of a dipole and two quadrupole magnets that focuses the particles' motion in the horizontal (x)-- and vertical (y)--directions \cite{BountisTompaidis}. After some appropriate scaling, the equations of the mapping are written as follows:
\begin{equation}
\label{4D}
\begin{cases}
x_{n+1}=2 c_x x_n - x_{n-1} - \rho x_n^2 +y_n^2\\
y_{n+1}=2 c_y y_n - y_{n-1}  +2 x_n y_n\\
\end{cases}
\end{equation}
where $\rho=\beta_x s_x/\beta_y s_y$, $c_{x,y}\equiv \cos{(2\pi q_{x,y})}$ and $s_{x,y}\equiv \sin{(2\pi q_{x,y})}$, $q_{x,y}$ is the so-called betatron frequencies and $\beta_{x,y}$ are the betatron functions of the accelerator. As in \cite{BountisKollmann}, we assume that $\beta_{x,y}$ are constant and equal to their mean values, i.e. proportional to $q_{x,y}^{-1}$ ($q_x=0.21$, $q_y=0.24$) and place our initial conditions at a particular point in $4$-dimensional space associated with weak diffusion phenomena in the $y$--direction. In particular, our $(x_0,x_1)=(-0.0049,-0.5329)$ coordinates are located within a thin chaotic layer surrounding the islands around a 5-order resonance in the $x_n,x_{n-1}$ plane of a purely horizontal beam, i.e. when $y_n=y_{n-1}=0$. We then place our initial $y_1,y_0$ coordinates very close to zero and observe the evolution of the $y_n$s indicating the growth of the beam in the vertical direction as the number of iterations $N$ grows.

\begin{figure}
\includegraphics[height=9.5cm,angle=0,clip=]{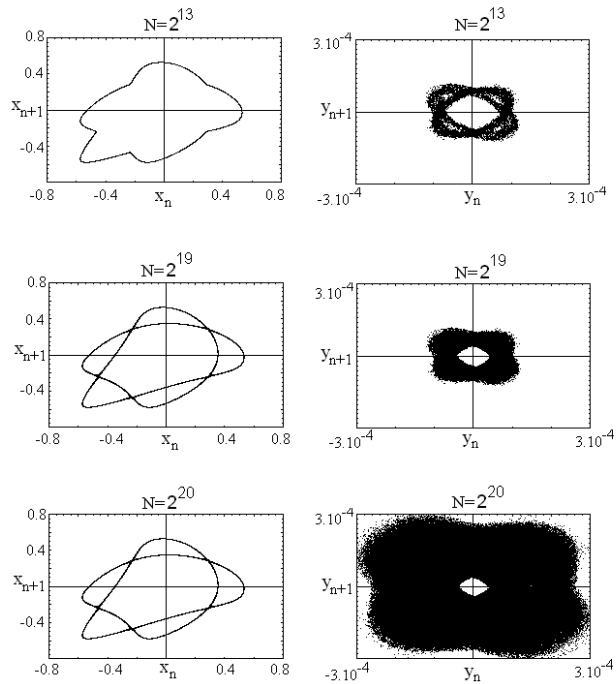}
\caption{The $x_n, x_{n+1}$ (first column of panels) and $y_n, y_{n+1}$ (second column of panels) projections of a chaotic orbit of \eqref{4D}, with $q_x=0.21$, $q_y=0.24$, $x_0=-0.0049$ and initial conditions $x_1=-0.5329$, $y_0=0.0001$ and $y_1=0$ (case II of Table III). $N$ represents the number of plotted iterates. \label{fig_12}}
\end{figure}

Let us observe this evolution in Fig.~\ref{fig_12} separately in the $x_{n+1},x_n$ (first column) and $y_{n+1},y_n$ (second column) $2$--dimensional projections of our chaotic orbits. Clearly the behavior of these projections is very different: In the $x$--plane the motion keeps evolving in a thin chaotic layer around five islands, ``feeding'' as it were the $(y_n,y_{n+1})$ oscillations, which show an evidently slow diffusive growth of their amplitude outward.

\begin{figure}
\includegraphics[height=12cm,angle=0,clip=]{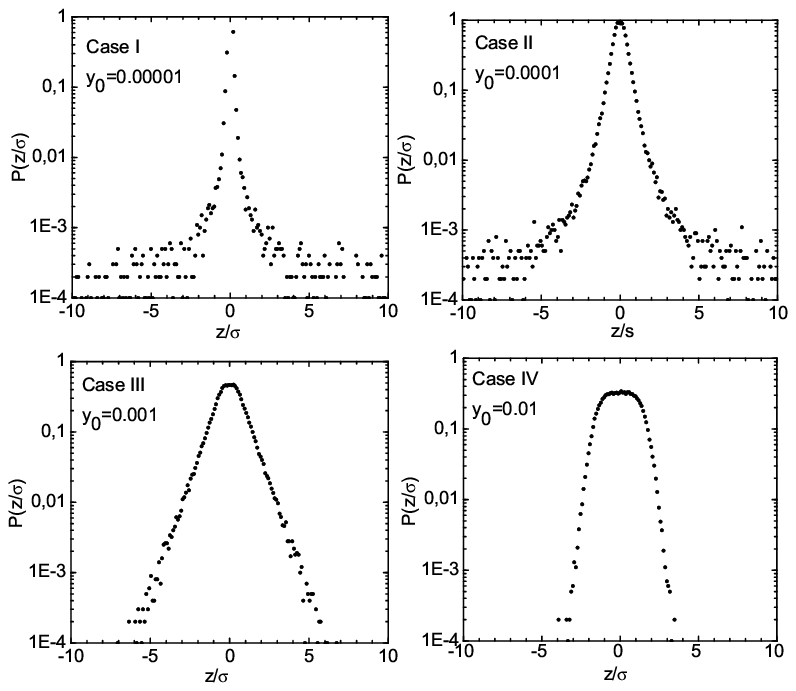}\vspace{-0.5cm}
\caption{Pdfs of the normalized sums of iterates of the $y$--chaotic orbit of the $4$--dimensional map \eqref{4D}, for different $y_0$. In all cases,  $q_x=0.21$, $q_y=0.24$, $x_0=-0.0049$, $x_1=-0.5329$ and $y_1=0$. The number of (summed) iterates is $N=2^{19}$, and the number of randomly chosen initial conditions within an interval $[0.9 y_0, y_0]$ is $N_{ic}=10^5$. \label{fig_13}}
\end{figure}

In Table~\ref{y0max} we list, for different initial values of $y_0$ ($y_1=0$), the maximum amplitude of the $y$--oscilaltions, $y_{max}$, while Fig.~\ref{fig_13} shows the corresponding pdfs of the normalized sums of iterates of the $y_n$-variable. Note that, just as in the case of $2$--dimensional maps, these distributions are initially $q$-{\it Gaussian}-like evolving slowly into {\it triangular}-like distributions, which finally approach Gaussians. In Fig.~\ref{fig_13} we follow this evolution by performing four computations of $N=2^{19}$ iterates starting with $y_0$ which increases every time by a factor of 10.

\begin{table}
\caption{Estimation of $(y_{max})$--coordinate after the diffusion process occurred along $N=10^6$ iterations, for different $y$--motion initial conditions $y_0$. In all cases,  $q_x=0.21$, $q_y=0.24$, $x_0=-0.0049$, $x_1=-0.5329$, and $y_1=0$.\label{y0max}}
 \begin{ruledtabular}
\begin{tabular}{ccc}
$\mbox{case} $ &$y_0$ &$y_{max}$  \\ \hline
I & $0.00001$ & $0.00002$ \\
II & $0.0001$ & $0.0003$ \\
III & $0.001$ & $0.004$ \\
IV & $0.01$ & $0.015$ \\
\end{tabular}
 \end{ruledtabular}
\end{table}

This similarity with the $2$--dimensional case makes us suspect that the orbits of our $4$--dimensional map also follow a sequence of weakly chaotic QSS, whose time--evolving features are depicted in plots of the $y$--motion in Fig~\ref{fig_12} (second column), for increasing $N$. Note, for example, that one such QSS with a maximum amplitude of about $0.00001$ is observed up to $N> N=2^{19}$, diffusing slowly in the $y$--direction. The pdf of this QSS is shown in the upper left panel of Fig~\ref{fig_13} and has the shape of a $q$--Gaussian up to this value of $N$. However, for higher values of the $y_0$ initial condition, due to the sudden increase of the $y_n$ amplitudes at $N=2^{20}$, the ``legs'' of the pdf are lifted upward and the distribution assumes a more triangular shape.

\begin{figure}\vspace{-2.5cm}
\includegraphics[height=12cm,angle=0,clip=]{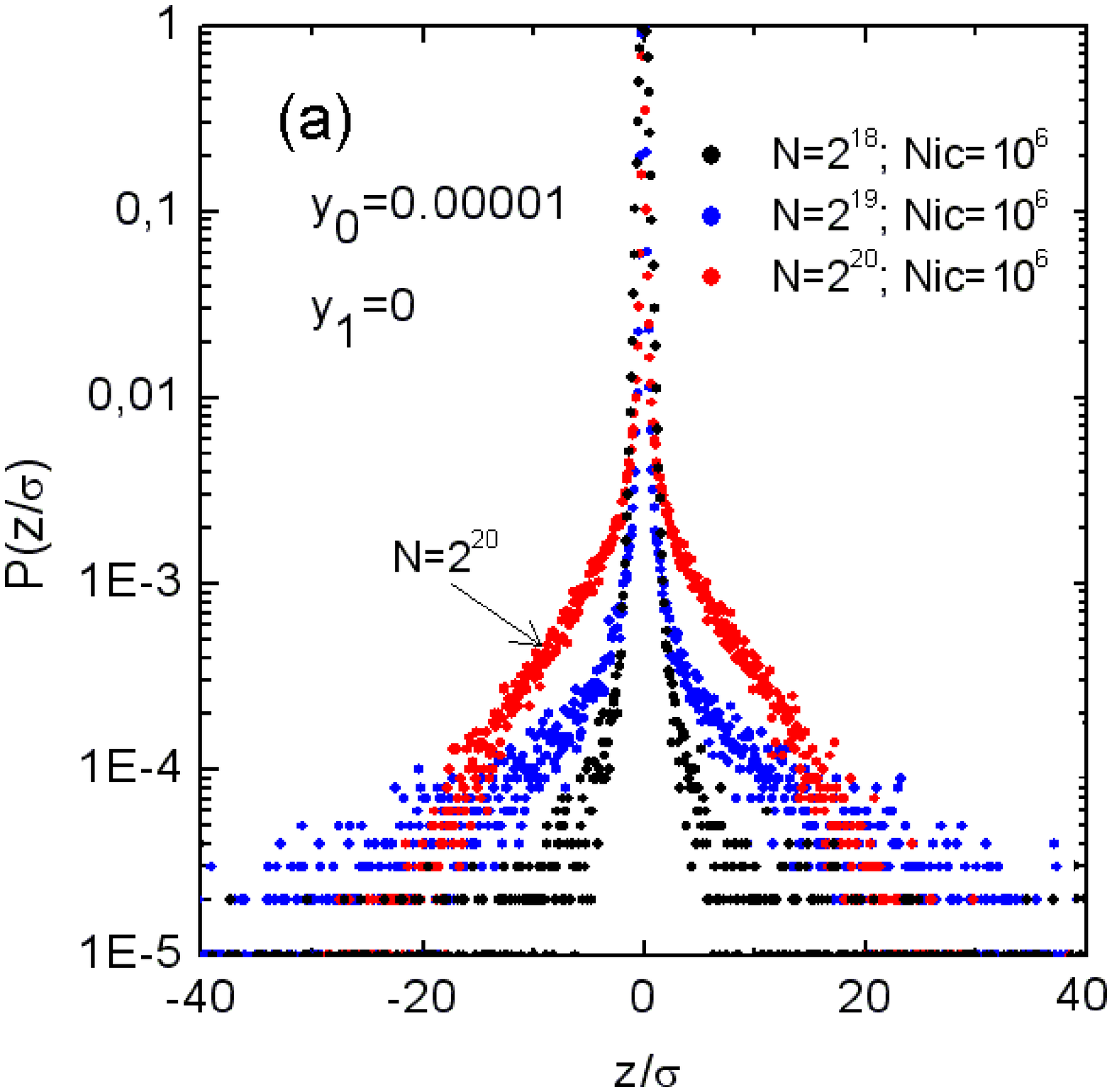}\vspace{-5cm}
\includegraphics[height=12cm,angle=0,clip=]{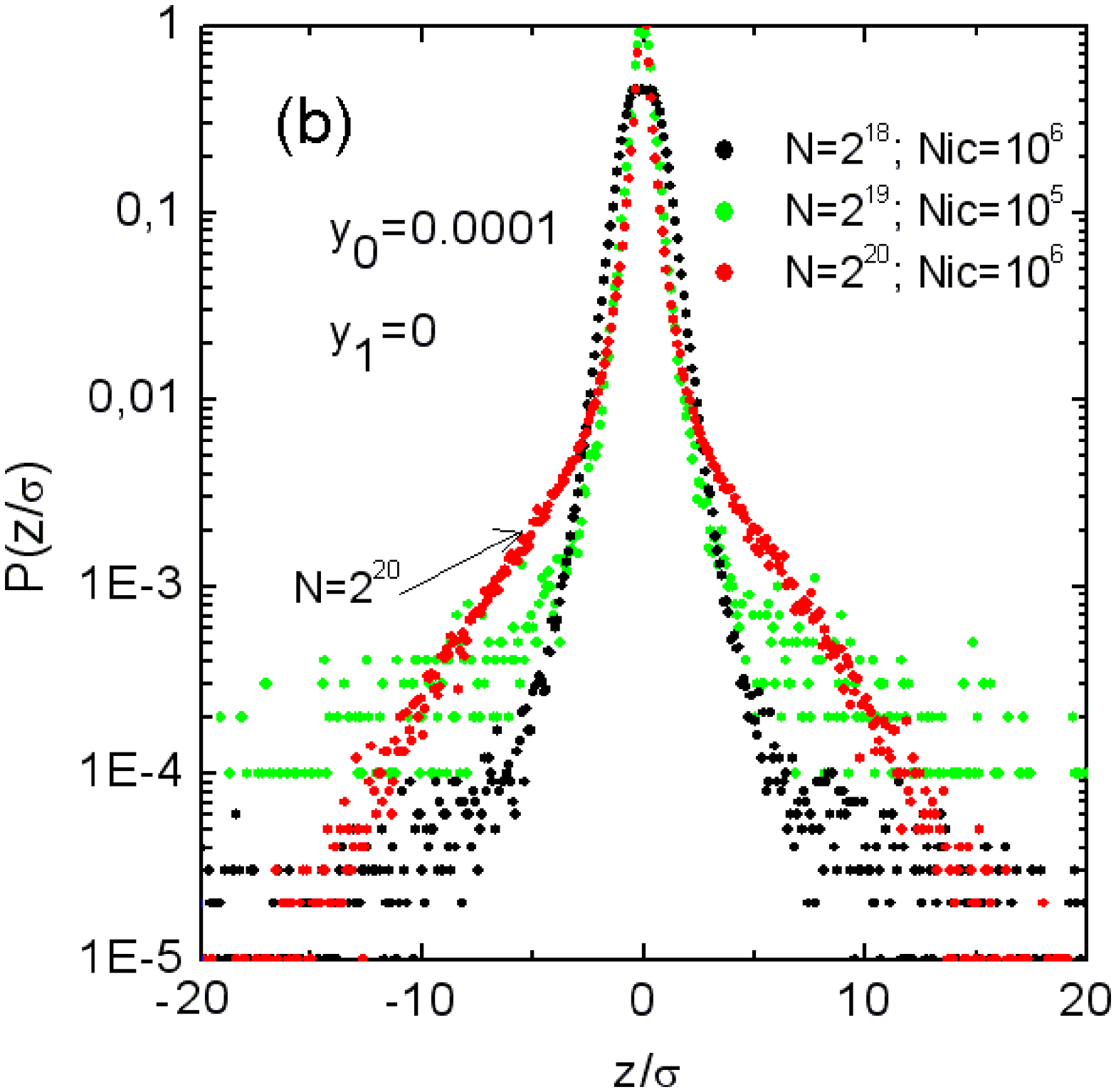}\vspace{2.5cm}
\caption{(Color online) Pdfs of the normalized sums of iterates of the $y$--chaotic orbit of the $4$--dimensional map, for different initial conditions $y_0$ and numbers of (summed) iterates $N$.  $N_{ic}$ is the number of randomly chosen initial conditions from an interval $[0.9 y_0, y_0]$. In all cases,  $q_x=0.21$, $q_y=0.24$, $x_0=-0.0049$, $x_1=-0.5329$, and $y_1=0$.  \label{fig_14}}
\end{figure}

This rise of the pdf ``legs'' to a triangular shape is shown in more detail in Fig.~\ref{fig_14}, for initial conditions $y_0=10^{-5},10^{-4}$, as the number of iterations grows to $N=2^{20}$. Clearly, the \textit{closer we start} to $y_0=y_1=0$ the more our pdf resembles a $q$--Gaussian, while as we move further out in the $y_0$--direction our pdfs tend more quickly towards a Gaussian--like shape. This sequence of distributions is reminiscent of what we found for the $2$--dimensional MacMillan map at $(\epsilon=0.9, \mu=1.6)$. Recall that, in that case also, a steady slow diffusion was observed radially outward, similar to what was observed for the $4$--dimensional map \eqref{4D}, which does not appear to be limited by a closed invariant curve in the $x_n,y_n$ plane.

One might wonder if it is possible to obtain for the $4$-dimensional map \eqref{4D} also long--lived $q$--Gaussian pdfs of the type we found in the
$2$--dimensional MacMillan map. The likelihood of this occurrence is small, however, as all orbits we computed for the accelerator map \eqref{4D} eventually \textit{escaped to infinity}! This implies that stickiness phenomena on island boundaries and sets of cantori are more dominant and tend to slow down diffusion more in the plane of 2--dimensional maps like the MacMllan map than the $4$--dimensional space of the accelerator map. It would, therefore, be very interesting to study, in a future paper, higher--dimensional maps whose chaotic orbits \textit{never escape} to infinity (e.g. coupled standard maps) and compare their statistics with what we have discovered for the examples treated in the present paper.

\section{Conclusions}

Our work serves to connect different types of statistical distributions of chaotic orbits (in the context of the Central Limit Theorem) with different aspects of dynamics in the phase space of conservative systems. What we have found, in several examples of the McMillan and Ikeda 2--dimensional area preserving maps as well as one case of a 4--dimensional symplectic accelerator map, is that $q$-Gaussians approximate well quasi-stationary states (QSS), which are surprisingly long--lived, especially when the orbits evolve in complicated chaotic domains surrounding many islands. This may be attributed to the fact that the maximal Lyapunov exponent in these regions is small and the dynamics occurs close to the so--called ``edge of chaos'' where stickiness effects are important near the boundaries of these islands.

On the other hand, in simpler--looking chaotic domains (surrounding e.g. only two major islands) the observed QSS passes, as time evolves, from a $q$--Gaussian to an exponential pdf and may in fact become Gaussian, as the number of iterations becomes arbitrarily large. Even in these cases, however, the successive QSS are particularly long-lasting, so that the Gaussians expected from uniformly ergodic motion are practically unobservable.

Interestingly enough, similar results have been obtained in N-dimensional Hamiltonian systems \cite{Antonopoulos,Leo2010} describing FPU particle chains near nonlinear normal modes which have just turned unstable as the total energy is increased. Since these models evolve in a multi--dimensional phase space, the $q$--Gaussian pdfs last for times typically of the order $10^6$, then pass quickly through the triangular stage and converge to Gaussians, as chaotic orbits move away from thin layers to wider ``seas'', where Lyapunov exponents are much larger. However, as long as the motion evolves near an ``edge of chaos'' region the distributions are $q$-shaped for long times, exactly as we found in the present paper.

These conclusions are closely related to results obtained by other authors \cite{Baldovin1,Baldovin2}, who also study QSS occurring in low-dimensional Hamiltonian systems like $2$-D and $4$-D maps, but not from the viewpoint of sum distributions. They define a variance of momentum distributions representing a temperature-like quantity $T(t)$ and show numerically that $T(t)$ follows a ``sigmoid'' curve starting from small values and converging to a final value, which they identify as the Boltzmann Gibbs (BG) state. Although their initial conditions are spread over a wide domain and do not start from a precise location in phase space as in our studies, they also discover many examples of QSS which remain at the initial temperature for very long times, before finally converging to the BG state.


\section*{ACKNOWLEDGMENTS}

One of us (T. B.) is grateful for the hospitality of the Centro Brasileiro de Pesquisas Fisicas, at Rio de Janeiro, during March 1 -- April 5, 2010, where part of the work reported here was carried out. We acknowledge partial financial support by CNPq, Capes and Faperj (Brazilian Agencies) and DGU-MEC (Spanish Ministry of Education) through Project PHB2007-0095-PC.



\end{document}